\documentstyle[12pt]{article}

\begin{document}
\begin{titlepage}

\begin{titlepage}
\hbox to \hsize{\hfil hep-th/9404346}
\hbox to \hsize{\hfil IHEP 94--53}
\hbox to \hsize{\hfil April, 1994}

\vfill
\large \bf
\begin{center}
 $B_c$--MESON PRODUCTION IN HADRON--HADRON COLLISIONS
\end{center}
\vskip 1cm
\normalsize
\begin{center}
{\bf S.R.~Slabospitsky \footnote{E-Mail: SLABOSPITSKY@mx.ihep.su }} \\
{\small Institute for High Energy Physics, 142284 Protvino, Moscow Region,
RUSSIA}
\end{center}
\vskip 2.cm
\begin{abstract}
\noindent
The $B_c$--meson production cross section was calculated in the
perturbative QCD. Various distributions of the charged leptons from the
 $B_c \to l \nu J/\psi(\to l^+ l^-)$ decay were obtained. The
$B_c$--meson mass measurement from such decays is also discussed.
\end{abstract}
\vfill
\end{titlepage}

\section{\bf Introduction}

The investigation of the properties and production mechanisms of
hadrons, composed of heavy quarks
($J/\psi$, $\Upsilon$, $\dots$), is one of the actual problems of
modern high energy physics.
Among such types of hadrons the most interesting are
$B_c(b \bar c)$--mesons -- yet not undiscovered mesons composed of
heavy quarks of different flavours~\cite{1}. Indeed, practically
all properties of the
$B_c$--family (the mass spectrum, decay widths, etc) are expected to be
calculable  by making use of
potential model~\cite{2,3} or QCD sum rules~\cite{4}
and there is no need to introduce new parameters. In such approach
all phenomenological parameters can be determined  form the description
of the $J/\psi$ and $\Upsilon$ families. On the other hand, due to the
fact that $B_c$--meson is a bound state of different flavoured heavy quarks
the study of the $B_c$ family spectroscopy  can be  used as a good method
to verify the potential model approach.

  Various $B_c$ weak decay properties were calculated
up to the present time (see, for example,~\cite{4,5}).
The $B_c$ lifetime computed by potential models is
$\tau_{B_c} \simeq 5 \cdot 10^{-13}$~s
(i.e. $\Gamma_{tot}(B_c) = 1.3 \cdot 10^{-3}$~eV \cite{2,3}).
The corresponding value from QCD sum rules is
$\tau_{B_c} = 9 \cdot 10^{-13}$~s
(i.e. $\Gamma_{tot}(B_c) = 7.4 \cdot 10^{-4}$~eV \cite{4}).
As a general characteristics, in the $B_c$ decays the presence of the
$J/\psi$ in the final states is rather frequent
($Br(B_c \to J/\psi X) = (20 \div 24)\%$ \cite{5}). This property is
explored in the current experimental searches
of the $B_c$--mesons~\cite{1}. The best decay channel that allows for full
reconstruction is $B_c \to J/\psi \pi^{\pm}$.
But due to the small probability of such decay
 ($Br(B_c \to J/\psi \pi^{\pm}) = (0.2\div0.4)\%$
\cite{5}), it would be rather difficult to search for the $B_c$--mesons
in this decay mode. A more preferable decay channel is
$B_c \to l \nu J/\psi$ with
$Br(B_c \to l \nu J/\psi) = (2.\div 4.)\%$ \cite{5}.
One could hope to establish the detection of a $B_c$ signal observing
the presence of a third lepton coming from the same secondary vertex
as the $J/\psi$. That requirement should significantly reduce possible
background.

In the present article the $B_c$--meson production cross section was
 calculated in the frame work of the perturbative QCD. The improvement
of the direct amplitude calculation method~\cite{6} was proposed.
This trick allowed one to write the  product of two Dirac spinors (describing
the fermion--antifermion system) in compact Lorentz--invariant form. The
decay $B_c \to l \nu J/\psi(\to l^+ l^-)$ was studied in detail.
Various charged lepton distributions were obtained. The measurement
of the $B_c$--meson mass from such decay was also discussed.

The article is organized as follows. The calculation of the amplitude
of the subprocesses of quark--qntiquark and gluon annihilation in the
perturbative QCD is described in Section~2.
The $B_c$--meson production cross section in hadron--hadron collisions
is presented in Section~3. In Section~4 there is discussed
the $B_c \to l \nu J/\psi(\to l^+ l^-)$ decay, charged lepton spectra from
this decay and the determination of $M_{B_c}$. The main results were
summarized in Conclusion. The basic relations needed for the
$B_c$ production amplitude evaluation are presented in
Appendixes~A and~B.

\section {\bf Amplitude calculation}

The  $B_c$--meson production in hadron--hadron collisions in the lowest
order of the perturbative QCD is described by the following subprocesses~:
\begin{eqnarray}
 q(r_1) \bar q(r_2) \to B_c(p) \bar b(k_1) c(k_2) \label{r1} \\
 g(r_1) g(r_2) \to B_c(p) \bar b(k_1) c(k_2) \label{r2}
\end{eqnarray}
where the 4--momenta of the corresponding particles are denoted in the
brackets. The $q \bar q$ (gluon) annihilation subprocess is described
by 7 (36) Feynman diagrams. Some diagrams are presented on Fig.1.

Since $b$-- and $c$--quarks producing $B_c$--meson are rather heavy
then one could treat them as lying near a mass shell~:
\begin{eqnarray}
 k_1^2 \simeq m_1^2 = m_b^2, \quad  k_2^2 \simeq m_2^2 = m_c^2.
\end{eqnarray}

In this article the lightest $B_c(0^-)$--meson production was investigated.
For this case, taking into account~(3) the transition vertex of $b$ and
$c$--quarks into $B_c(0^-)$ (Fig.2) can be written as follows~\cite{krt}~:
\begin{eqnarray}
 G(-k_2) \Psi_{B_c}(p) G(k_1) \simeq g_B (m_2-k_2) \gamma^5
(m_1 + k_1) \frac{\delta_{ij}}{\sqrt{3}}
  = \frac{f_{B_c}}{12} \gamma^5(M + \hat p) \delta_{ij}. \label{gbc}
\end{eqnarray}
where $G(k)$ is the quark propagator; the $M(p)$ is the mass (momentum)
of $B_c$--meson.

To derive relation (\ref{gbc}) one uses the fact that due to (3) the momenta
of $b$ and $c$--quarks (produced the $B_c$) are equal~:
\begin{eqnarray}
 k_1 = \frac{m_1}{M} p, \quad  k_2 = \frac{m_2}{M} p. \nonumber
\end{eqnarray}

The effective weak decay constant $f_{B_c}$ (see (\ref{gbc})) is related to the
nonrelativistic wave function and $B_c$--meson weak decay width in the
following way~\cite{2,3,5}~:
\begin{eqnarray}
 f_{B_c} &=& \sqrt{\frac{12}{M}} \Psi(0) \simeq 510 \, \, \mbox{MeV} \\
\Gamma(B_c \to \mu \nu) &=& \frac{G_F^2}{8 \pi} |V_{bc}|^2
 f^2_{B_c} M m_{\mu}^2 (1- \frac{m^2_{\mu}}{M^2})^2 \nonumber.
\end{eqnarray}

The axial gauge~\cite{t1} was used for the calculation of the gluon propagators
and polarization density matrix of a gluons. For a gauge vector it is
suitable to use the $B_c$--meson momentum~$p$. Then the gluon propagator
has the form~:
\begin{eqnarray}
 D^{\alpha \beta}(q) = \frac{1}{q^2} ( g^{\alpha \beta}
 - \frac{p^{\alpha} q^{\beta} + p^{\beta} q^{\alpha} }{(pq)}
 + \frac{M^2 q^{\alpha} q^{\beta}} {(pq)^2}). \label{prop}
\end{eqnarray}

The real gluon polarization vectors $\epsilon_i^{\mu}(i=1,2)$
are defined as follows~:
\begin{eqnarray}
 \epsilon_i^{\mu} = e_i^{\mu} - \frac{(pe_i)}{(pr)}r^{\mu}, \quad
 (\epsilon_i^{\mu} \epsilon_j^{\mu}) = -\delta^{ij}, \quad
 (\epsilon_i  r) = (\epsilon_i  p) = 0, \nonumber
\end{eqnarray}
here $r$ is gluon momentum.

 The vectors $e_1$ and $e_2$ are ($r_{\bot}$ is transverse momentum of
a gluon)~:
\begin{eqnarray}
 &a)& r_{\bot} = 0 \quad e_1 = (0,0,1,0), \quad e_2 = (0,1,0,0) \\
 &b)& r_{\bot} > 0 \quad e_1 = \frac{1}{r_{\bot}}(0,-r_2,r_1,0),
 \quad e_2 = \frac{1}{r_{\bot} |\vec r|}(0,r_1 r_3, r_2 r_3, -r^2_{\bot}).
 \nonumber
\end{eqnarray}

Since the number of diagrams describing subprocesses~(\ref{r1})
and (\ref{r2}) is relatively large
(7 diagrams for $ q \bar q \to \dots $ and 36 ones for $ g g \to \dots $),
 it is natural to evaluate the
amplitude instead of the matrix element squared. The particles participating
in subprocesses~(\ref{r1}) and (\ref{r2}) are unpolarized. In this case
for the calculation of the amplitude one can write the product of two Dirac
spinors in the compact Lorentz--invariant form
 (see Appendix~A for detailed derivation)~:
\begin{eqnarray}
 u(-k_1) \bar u(k_2) &\Rightarrow& \frac{1}{2 \sqrt{2 m_1 m_2}}
 (m_1-\hat k_1) (1+\gamma^5)\gamma^{\xi} (m_2 + \hat k_2) \label{m1} \\
 u(-r_1) \bar u(r_2) &\Rightarrow&
\frac{\hat r_1(1-\gamma^0) (1+\gamma^5)\gamma^{\xi}
(1+\gamma^0) \hat r_2 } {4 \sqrt{2  E_1 E_2}} \label{m2}
\end{eqnarray}
where $E_{1,2}$ are the energies of the initial massless quarks.

The vector index $\xi$, which appeared in the right hand of
relations~(\ref{m1}) and (\ref{m2}) corresponds to four possible
combinations of the helicities of two fermion system~\cite{6}.

Taking into account relations~(\ref{m1}) and (\ref{m2}) the
amplitude of subprocess~(\ref{r1}) can be written as a product
of two traces (corresponding to two fermion lines~: one trace
 is due to initial $q$--quarks, and the second one is due to
final $\bar b c$ quarks)~:
\begin{eqnarray}
 M(q \bar q \to B_c \bar b c) \sim  T_q^{\eta \xi} =
 \frac{ Tr[\hat r_1 (1-\gamma^0) (1+\gamma^5)\gamma^{\eta}
(1+\gamma^0) \hat r_2 \dots] } {4 \sqrt{2  E_1 E_2}} \nonumber \\
 \frac{Tr[(m_1-\hat k_1) (1+\gamma^5)\gamma^{\xi} (m_2 + \hat k_2)
\dots ]} {2 \sqrt{2 m_1 m_2}}. \label{mq}
\end{eqnarray}
The gluon annihilation amplitude (subprocess (\ref{r2})) is
proportional to one trace~:
\begin{eqnarray}
 M(g g \to B_c \bar b c) \sim T_g^{\xi} =
 \frac{Tr[(m_1-\hat k_1) (1+\gamma^5)\gamma^{\xi} (m_2 + \hat k_2)
\dots ]} {2 \sqrt{2 m_1 m_2}}. \label {mg}
\end{eqnarray}

With relations  (\ref{mq}) and (\ref{mg}) it is a trivial task to
calculate the matrix elements squared for subprocesses
(\ref{r1}) and (\ref{r2})~:
\begin{eqnarray}
 |M(q \bar q \to B_c \bar b c)|^2 &=& T_q^{\eta \xi} T_q^{\eta \xi}, \\
 |M(g g \to B_c \bar b c)|^2 &=& T_g^{\xi} T_g^{\xi},
\end{eqnarray}
where the sum over $\eta$ and $\xi$ indexes is performed as usual~:
\begin{eqnarray}
\mbox{e.g.} \quad T_g^{\xi} T_g^{\xi} = T_g^0 T_g^0 -
  T_g^1 T_g^1 - T_g^2 T_g^2 - T_g^3 T_g^3.
\end{eqnarray}

The full expressions for the amplitude for subprocesses~(\ref{r1}) and
 (\ref{r2}) are rather bulky  and we do not write them here. But we present
all expressions needed to evaluate the amplitudes in
Appendixes~A and~B.

\section {\bf $B_c$--meson production cross section}

The $B_c$--meson production cross section in the parton model has the form~:
\begin{eqnarray}
\sigma(hh \to B_c X) = \sum \limits_{i,j} \int
d x_1 d x_2 f^{(1)}_i(x_1,Q^2) f^{(2)}_i(x_2,Q^2)
 \hat \sigma(i j \to B_c \bar b c),
\end{eqnarray}
where the sum is performed over all partons, participating in
subprocesses~(\ref{r1}) and (\ref{r2}) of $B_c$ production. We
use the CTEQ collaboration parametrization of the structure functions
$f(x,Q^2)$ from \cite{stf}. The evolution parameter
 $\sqrt{Q^2}$ and particle masses used are as follows~:
\begin{eqnarray}
 \sqrt{Q^2} &=& \sqrt{\hat s}, \nonumber \\
 M& =& m_{B_c} = 6.3 \, \mbox{GeV},\, \,
 m_1 = m_b = 4.8 \, \mbox{GeV}, \, \, m_2 = m_c = 1.5 \, \mbox{GeV}.
\label{par}
\end{eqnarray}
where $\hat s$ is the total energy squared of the partons from
subprocess (\ref{r1}) and (\ref{r2}).

At the given parameter values from (\ref{par}) the inclusive
$B_c$--meson total production cross section at the $FNAL$ collider  energies
($\sqrt{s} = 1.8$~TeV, $p \bar p$--collisions) equals~:
\begin{eqnarray}
 \sigma_0(p \bar p \to (B^+_c + B_c^-) X) =  15.2 \, \, \mbox{nb}, \label{s0}
\end{eqnarray}
This value could be compared with the latest calculations of the
$b \bar b$ production cross section at the same energies \cite{rey}~:
\begin{eqnarray}
 \sigma(p \bar p \to b \bar b X) =  (3.8 \div 5.6) \cdot 10^4 \, \, \mbox{nb},
\nonumber
\end{eqnarray}

One should note that the cross section due to $q \bar q$--annihilation
is negligibly small ($\leq 1\%$) as compared to gluon annihilation
subprocess and we do not take it into account in what follows.

We investigated the dependence of the $B_c$  production cross section on
(\ref{s0}) as a function of the evolution parameter
and masses of $b$-- and $c$--quarks. We used three variants for
$\sqrt{Q^2} = \sqrt{\hat s}$, $\sqrt{\hat s}/2$ and
$\sqrt{Q^2} = m_b+m_c+M_{B_c} = 2M_{B_c}$. Two variants
($\sqrt{\hat s}$ and $2M_{B_c}$) cover all the possible choice for
$\sqrt{Q^2}$. The range for $m_b = (4.5\div5.2)$~GeV (and
$m_c = M_{B_c} - m_b)$ covers all possible values of $b$--quark
mass used in the calculations.
The results are presented in Table~1. As one can see form this table
that under reasonable variation of the parameters the cross section
value change not more than 4~times~:
\begin{eqnarray}
 \sigma_0(p \bar p \to (B^+_c + B_c^-) X) = (12.\div 45.) \, \, \mbox{nb},
 \label{st0}
\end{eqnarray}

The additional sources for $B_c(0^-)$--mesons are the production of the
excited states $B_c^{\ast}$--mesons. These states $B_c^{\ast}$
(in the contrast to $\psi', \cdots$, $\Upsilon' \cdots$) transit into
$B_c(0^-)$ with unit probability. The production
of the $B_c^{\ast}$ leads to the rise of the total
$B_c$--meson production cross section up to  ($5\div6$) times
(see  \cite{ch}, \cite{kit}).  Thus
total $B_c$--meson production cross section, calculated in the
perturbative QCD could be estimated as follows~:
\begin{eqnarray}
 \sigma_{tot}(p \bar p \to B_c X) = \sum \limits_{B_c{\ast}}
 (1\div4)\sigma(p \bar p \to B_c^{\ast} X) \equiv (1\div 4) 5
 \sigma_0(p \bar p \to B_c X) \nonumber \\
 = (60. \div 225.) \, \mbox{nb}, \label{st}
\end{eqnarray}
where the uncertainty $(60 \div 225)$~nb reflects the variation of the
parameters $\sqrt{Q^2}$, $m_b$ and  $m_c$.

Resulted estimates (\ref{st}) allowed one to obtain the following relation
($\sqrt{s} = 1.8$~TeV)~:
\begin{eqnarray}
 R_{B_c} = \frac{\sigma_{tot}(p \bar p \to B_c X)}
 {\sigma(p \bar p \to b \bar b X)} = (1.1\div 5.9) \cdot 10^{-3}. \label{rel}
\end{eqnarray}

We note that obtained value (\ref{rel})  confirms the rough estimate
 ($R_{B_c} = 3 \cdot 10^{-3}$), received early in~\cite{vk}, and
coincides with results of the $HERWIG$ Monte--Carlo program simulations
($R_{B_c} = (1.3\pm.2) \cdot 10^{-3}$ \cite{herw}).
At the same time the obtained value of cross section (\ref{st})
is essentially higher than the results of article~\cite{ch}.

The total $B_c$--meson production cross section behaviour versus
 $\sqrt{s}$ is presented on Fig.3. The differential distributions
of $B_c$--mesons, $b$-- and $c$--quarks over pseudorapidity
and transverse momentum are presented on Fig.4.

\section {\bf Search for $B_c$--meson in the decay channel
$B_c \to l \nu J/\psi (\to l^+ l^-)$}

In this section we consider the $B_c$--meson decay~:
\begin{eqnarray}
 B_c(p) \to l(q_1) \nu(q_2) J/\psi(k) (\to l^+(q_3) l^-(q_4)) \label{22}
\end{eqnarray}
where the momenta of the particle are denoted in the brackets.

Under on--shell $J/\psi$--meson  assumption the amplitude of this
decay can be written as follows~\cite{yf}~:
\begin{eqnarray}
 A = \frac{G_F}{\sqrt{2}} V_{bc} L^{\nu} H^{\nu} \nonumber
\end{eqnarray}
where the $G_F$ is a Fermi constant; $V_{bc}$ is an element of the
Kobayashi--Maskawa matrix;
$L^{\nu} = \bar \mu(q_1) \gamma^{\nu}(1 - \gamma^5) \nu(q_2)$
is  lepton current.  The $H^{\nu}$ is a matrix element of the hadronic
current $J^{\nu} = b \gamma^{\nu}(1 - \gamma^5) \bar c$. In the framework
of the constituent quark model $H^{\nu}$ has the form~\cite{yf}~:
\begin{eqnarray}
 H^{\nu} = \frac{2}{(M+m_{\psi})} [a_1(t) g^{\nu \sigma}
 + a_2(t) Q^{\sigma}k^{\nu}
 + i a_3(t)\epsilon^{\nu \sigma \alpha \beta}Q^{\alpha} k^{\beta})
\varepsilon^{\sigma},
\end{eqnarray}
where $M$ and $m_{\psi}$ are the $B_c$ and $J/\psi$ masses;
$Q = q_1 + q_2$; $k$ and $\varepsilon$ are momentum and polarization
vector of the $J/\psi$--meson; $t = (q_1+q_2)^2$.

The transition form--factors $a_{1,2}(t)$ are proportional
to $a_3(t)$~\cite{yf}~:
\begin{eqnarray}
 a_1(t) = - \frac{(M^2 + m^2_{\psi} - t)}{2} a_3(t), \quad
 a_2(t) =  ( 1 -  \frac{m_{\psi}}{M}) a_3(t). \nonumber
\end{eqnarray}
The exact expression for $a_3(t)$ (see \cite{yf}) in the allowed
kinematical region can be parameterized as follows~:
\begin{eqnarray}
 a_3(t) = \frac{V(0)}{1 - t/m_k^2}, \nonumber
\end{eqnarray}
where $V(0) = 1.31$ and $m_k = 6.3$~GeV (for details see~\cite{yf}).

In the zero width of the $J/\psi$--meson approximation the width
of the $B_c$--meson decay~(\ref{22}) has the form~:
\begin{eqnarray}
\Gamma_{B_c} = \frac{96 (2\pi)^4 G_F^2 |V_{bc}|^2 Br}
 {M m^2 (M+m_{\psi})^2} \chi R_3(p - q_1 - q_2 - k) R_2(k - q_3 - q_4)
 \label{28}
\end{eqnarray}
where $Br = Br(J/\psi \to l^+ l^-)=0.07\%$ \cite{pdg} is
of the $J/\psi \to l^+ l^-$ decay probability; the phase space volume $R_n$
equals~:
\begin{eqnarray}
R_n(p - \sum_{i=1}^{n}q_i) = \delta^{(4)}(p - \sum_{i=1}^{n}q_i)
 \prod_{i=1}^{n} \frac{d^3 q_i}{(2\pi)^3 2 q_{i0}}. \nonumber
\end{eqnarray}

The matrix element squared $\chi$ from (\ref{28}) has the form :
\begin{eqnarray}
\chi = a_1^2 u_1 + a_2^2 u_2 + a_3^2 u_3 + a_1 a_2 u_4
+ a_1 a_3 u_5   + a_2 a_3 u_6, \label{x}
\end{eqnarray}
where
\begin{eqnarray*}
u_1 &=& 2[(q_1q_3)(q_2q_4) + (q_1q_4)(q_2q_3)], \\
u_2 &=& [2(q_1Q)(q_4Q)- \frac{m_{\psi}^2t}{2}][2(q_1k)(q_2k)
- \frac{m_{\psi}^2t}{2}], \\
u_3 &=& \frac{m_{\psi}^2t}{2}[(q_1q_3)^2 + (q_1q_4)^2 + (q_2q_3)^2
 + (q_2q_4)^2
 + 2(q_1q_3)(q_2q_4) + 2(q_1q_4)(q_2q_3)] \\
 &&- \frac{m_{\psi}^4t^2}{4} - 4[(q_1q_3)(q_2q_4) - (q_1q_4)(q_2q_3)]^2, \\
u_4 &=& -\frac{m_{\psi}^2t}{2}(kQ) + 2(q_1k)[(q_2q_4)(q_3Q)
+ (q_2q_3)(q_4Q)] \\
 && + 2(q_2k)[(q_1q_4)(q_3Q) + (q_1q_3)(q_4Q)], \\
u_5 &=& \frac{m_{\psi}^2t}{2}[(q_1k)-(q_2k)]
+ 2[(q_3Q)-(q_4Q)][(q_1q_3)(q_2q_4) - (q_1q_4)(q_2q_3)], \\
u_6 &=& 2[(q_3Q)-(q_4Q)] [
\frac{m_{\psi}^2t}{4}((q_1q_4) + (q_2q_3) - (q_1q_3) - (q_2q_4)) \\
 &&+ (kQ)((q_1q_3)(q_2q_4) - (q_1q_4)(q_2q_3))].
\end{eqnarray*}

The  decay width (\ref{22}) of the $B_c$--meson calculated with (\ref{28})
equals~:
\begin{eqnarray}
\Gamma(B_c \to l \nu J/\psi) = 41. \cdot 10^{-6} \, \mbox{eV}, \quad
\mbox{i.e.} \, \,
 Br(B_c \to l \nu J/\psi) = 2.\div 4.\% \quad \cite{5}. \label{br}
\end{eqnarray}

The distributions over the transverse momentum and pseudorapidity
of the charged leptons from decay (\ref{22}) in the reaction
(at $\sqrt{s} = 1.8$~TeV)
\begin{eqnarray}
 p \bar p \to B_c( \to 3 \mu^{\pm} \nu) X, \label{sig}
\end{eqnarray}
are presented on Fig.5. One should note that charged leptons are
produced with small $p_{\top}$. The behaviour of the charged lepton
integrated cross section (\ref{sig}) as a function of the
minimum transverse momentum $p_{\top \, l}$~:
\begin{eqnarray}
R_{\top}(p_{\top min}) \equiv
 \frac{
 \int_{p_{\top min}}
  \frac{d \sigma (p \bar p \to B_c(\to 3 \mu^{\pm}) X)}{d p_{\top l}} }
 { \int_{0}
  \frac{d \sigma (p \bar p \to B_c(\to 3 \mu^{\pm}) X)}{d p_{\top l}} }
  \label{rl}
\end{eqnarray}
are presented on Fig.6. As one can see from this figure, this ratio decrease
very rapidly with increasing $p_{\top min}$. We take into account the
experimental cuts used at $D\emptyset$ detectors at $FNAL$~\cite{d0}~:
\begin{eqnarray}
p_{\top l } \geq 3 \, \, \mbox{GeV}, \quad |\eta_l| \leq 3.3.
\end{eqnarray}
Then taking into account the estimate of the $B_c$--meson production
cross section~(\ref{st}), the behaviour~$R_{\top}$
and the value from~(\ref{br}) one can obtain~:
\begin{eqnarray}
 \sigma_0 (p \bar p \to B_c(\to 3 \mu^{\pm}) X)_{cuts} =
(0.6 \div 4.5) \, \, \mbox{pb}, \nonumber \\
 \sigma_{tot} (p \bar p \to B_c(\to 3 \mu^{\pm}) X)_{cuts} =
(3.2 \div 24.) \, \, \mbox{pb}. \label{sle}
\end{eqnarray}
These values correspond to the following number of events (at the
total luminosity $\int {\it L} dt$ = 20~pb$^{-1}$)~:
\begin{eqnarray}
 N_0 (p \bar p \to B_c(\to 3 l^{\pm}) X)_{cuts} &=& (12. \div 90.)
 \nonumber \\
 N_{tot} (p \bar p \to B_c(\to 3 l^{\pm}) X)_{cuts} &=& (64. \div 480.),
\label{nle}
\end{eqnarray}
where $\sigma_0(N_0)$ and $\sigma_{tot}(N_{tot})$ correspond to the
$B_c$--meson production cross sections (the number of events)
calculated in the perturbative QCD and with contribution from
$B_c^{\ast}$, respectively.

The obtained estimates demonstrate that the observation of the $B_c$ in
reaction~(\ref{sig}) at the $FNAL$ energy in
the $B_c \to 3 \mu^{\pm} X$ decay channel is quite a real task.

We make the following remark at that point. Estimates~(\ref{sle})
and  (\ref{nle}) should be treated as minimal. Indeed we considered
the $B_c$ product in the {\bf \it lowest} tree level of perturbative
QCD. Higher order subprocesses (such as
$g g \to g B_c \bar b c$) could result in significant increase
of the $B_c$ production cross section at high $p_{\top}$ and of
estimates~(\ref{sle}) and (\ref{nle}).

Now we proceed to the discussion of the possibility of a determine
the $B_c$--meson mass in decay channel~(\ref{22}).
The information about the $M_{B_c}$ can be obtained from the
consideration of various distributions of charged leptons.
Let us define the following variables~:
\begin{eqnarray}
 M_{l \psi} = \sqrt{(p_l + p_{\psi})^2} = \sqrt{(q_1 + k)^2}, \, \,
 M_{l \mu} = \sqrt{(p_{l^{\pm}} + p_{\mu})^2} = \sqrt{(q_1 + q_{3,4})^2}.
\end{eqnarray}
It is useful to compare the distributions over $M_{l \psi}$
and $M_{l \mu}$, calculated with matrix element squared $\chi$
 from (\ref{x}), with the same ones,
but calculated under assumption of the constant of the
matrix element squared (i.e. with pure phase space)~:
\begin{eqnarray}
 &&\frac{d\Gamma_{B_c}^{const}}{d M_{l \psi}} \sim
 \frac{(M^2_{B_c} - M_{l \psi}^2)(M_{l \psi}^2-m_{\psi}^2)}{M_{l \psi}},
\nonumber \\
 && \frac{d\Gamma_{B_c}^{const}}{d M_{l \mu}} \sim M_{l \mu}
 [ M^2_{B_c} \ln(\frac{M^2_{B_c}}{m_{\psi}^2 + M_{l \mu}^2})
- M^2_{B_c} +  m_{\psi}^2 + M_{l \mu}^2]. \nonumber
\end{eqnarray}

The corresponding distributions $d\Gamma_{B_c}/dM$ (normalized for the unit)
are presented on Fig.7 ($d\Gamma_{B_c}/dM_{l \mu}$) and
on Fig.8 ($d\Gamma_{B_c}/dM_{l \psi}$). As one can see from these
figures the matrix element is essentially not a constant. So the study
of the distributions over $M_{l \psi, \, l \mu}$ provide the determination
of the $M_{B_c}$, but with rather poor accuracy (due to the theoretical
uncertainties in the $\chi$~\cite{yf} and especially for the case of the small
number of events).

We propose for consideration the distribution over another variable, which
could allow for the determination of $B_c$ mass with better accuracy.
Let us consider a possibility when one can measure the direction of
the $3$--momentum of $B_c$--meson. That can be done by measuring
the interaction point of the $p \bar p$--collision and position of the $B_c$
decay vertex. In this case one can investigate the distribution
over another variable~$M_{\bot}$~:
\begin{eqnarray}
 M_{\bot} = q_{\bot} + \sqrt{q_{\bot}^2 + M_{l \psi}^2}, \quad
 m_{\psi} \leq  M_{\bot} \leq M,
\end{eqnarray}
where $q_{\bot} = (p_l + p_{\psi})_{\bot}$ is the transverse momentum of the
(lepton $+ \quad J/\psi$)--system with respect to the direction of the
$B_c$--meson momentum.

Under matrix element constant assumption the distribution
over this variable has the form~:
\begin{eqnarray}
 \frac{d\Gamma_{B_c}^{const} }{dM_{\bot} } &\sim&
 \frac{f( M_{\bot} / M_{B_c})}{M_{\bot}^2 \sqrt{M_{B_c}^2-M_{\bot}^2} },
\label{34} \\
f(z) &=& z^2(z^2-1/2)[Asin(z)-Asin(\frac{\alpha^2}{z})]
 + \frac{\alpha^2}{2}\sqrt{z^2-\alpha^4}  \nonumber \\
&&+ z(\frac{z^2}{2} - \alpha^2)\sqrt{1-z^2}
- \alpha^2 z^3 \ln[\frac{z(z+\sqrt{z^2-\alpha^4}}{\alpha^2(1+\sqrt{1-z^2}},
\nonumber
\end{eqnarray}
where $\alpha = m_{J/\psi} / M_{B_c}$ and $\alpha \leq z \leq 1$.

The distributions $d\Gamma_{B_c}/d M_{\bot}$ and
$d\Gamma_{B_c}^{const}/d M_{\bot}$ (normalized for unit)
are presented on Fig.9. As one can see from this figure these
distributions practically coincide (in contrast with the
distributions over $M_{l \psi}$ and $M_{l \mu}$).
Thus the theoretical uncertainties connected with $\chi$ are very small.
Moreover, as it follows from (\ref{34})
the distribution over this variable has a sharp kinematical peak at
$M_{\bot} = M_{B_c}$. So the study of the distribution over this
variable $M_{\bot}$ could be used as a good method for the determination
of the $B_c$--meson  mass.

\section*{Conclusion}

In this article the $B_c(0^-)$--meson production cross section was
calculated in the framework of the perturbative QCD. The obtained
$B_c$--meson production cross section
(with contributions from the excited $B_c^{\ast}$ states) is
about $\sim (2\div8) \cdot 10^{-3}$ of the total $b \bar b$ pairs
production cross section at $\sqrt{s} \geq 0.1$~TeV. This result is in
an good agreement with early results.

The possibility of the $B_c$--mesons  search in the
 $B_c \to l \nu J/\psi(\to l^+ l^-)$ decay channel was investigated in
details. The various charged lepton distributions from this decay were
obtained also. The possibility of the determination of the $B_c$ mass
in this decay was considered.

In conclusion the author expresses his gratitude to Yu.Antipov, D.Denisov,
 A.Efimov, V.Kisilev, M.Mangano, C.Quigg, A.Razumov, R.Rogalyov
and A.Volkov for stimulating discussions.

\vspace{1.5cm}
\hbox to \hsize {\hfill \it Received 18 April 1994}

\newpage
\appendix
\makeatletter
  \@addtoreset{equation}{section}
  \def\thesection{\Alph{section}}
  \def\theequation{\thesection.\arabic{equation}}
\makeatother

\section {}

Relations (\ref{m1}) and (\ref{m2})
can be obtained from the Fierz identity~\cite{t1}~:
\begin{eqnarray}
 [\gamma^{\xi}(1-\gamma^5)]_{ij} \, \, [\gamma^{\xi}(1+\gamma^5)]_{kl} \,
 =  2 (1+\gamma^5)_{il} \, \, (1-\gamma^5)_{kj}   \label{p1}
\end{eqnarray}
where $i,j,k,l$ are the matrix indexes. Here and below
repeated indexes imply the summation.

Let us discuss a process with two final unpolarized antifermion
$f_1(m_1, k_1)$ and fermion $f_2(m_2, k_2)$ ($m$ and $k$ are the mass and
$4$--momentum of the fermion, respectively).

Let $M = \bar u(k_2) A u(-k_1)$ and $M' = \bar u(k_2) A' u(-k_1)$
be the two parts of the amplitude, connected
with the fermion line (for the total amplitude $M = M'$).
Then  after the summation
over the polarizations of the fermions one obtains the following term
of the amplitude squared~\cite{t1}~:
\begin{eqnarray}
 M'^{+} M  = \Sigma_{\rm{pol}} \bar u(-k_1) \bar A' u(k_2) \bar u(k_2)
 A u(-k_1) = Tr(m_1-\hat k_1)\bar A' (m_2+\hat k_2) A    \label{p2}
\end{eqnarray}
where $\bar A' = \gamma^0 A'^+ \gamma^0$.

\noindent Take into account that the fermions $f_1$ and $f_2$
are on--shell~:
\begin{eqnarray}
 (m \pm \hat k) = \frac{1}{2m} (m \pm \hat k)  (m \pm \hat k)
 = \frac{1}{2m} (m \pm \hat k) (1\pm \gamma^5) (m \pm \hat k)  \label{p3}
\end{eqnarray}
Substituting (\ref{p3}) into (\ref{p2}) and bearing in mind (\ref{p1})
one gets~:
\begin{eqnarray}
 M'^{+} M  = \frac{1}{8 m_1 m_2}
 &&Tr[(m_2+\hat k_2) \gamma^{\xi} (1+\gamma^5) (m_1 - \hat k_1) \bar A']
 \nonumber \\
 \cdot &&Tr [ (m_1 - \hat k_1) \gamma^{\xi} (1-\gamma^5) (m_2 + \hat k_2) A].
\label{p4}
\end{eqnarray}
Let us remember that the trace of $\gamma$--matrixes product
does not change when the order of the matrixes is reversed~\cite{t1}.
 Then from~(\ref{p4}) we obtain~:
\begin{eqnarray}
 M'^{+} M  = \frac{1}{8 m_1 m_2}
&&Tr [ (m_1 - \hat k_1)  (1+\gamma^5) \gamma^{\xi} (m_2 + \hat k_2) A]
 \nonumber \\
\cdot &&Tr [ (m_1 - \hat k_1) (1+\gamma^5) \gamma^{\xi} (m_2 + \hat k_2) A'].
\end{eqnarray}
Thus we come to relation (\ref{m1})~:
\begin{eqnarray}
 u(-k_1) \bar u(k_2) \Rightarrow \frac{1}{2 \sqrt{2 m_1 m_2}}
 (m_1-\hat k_1) (1+\gamma^5)\gamma^{\xi} (m_2 + \hat k_2) \label{p5}
\end{eqnarray}
One should note that the appearance of an additional index~$\xi$ in
(\ref{p5})  is quite natural. This index corresponds to four possible
polarization states of a two fermion system.

The obtained relation (\ref{p5}) can be applied to the case of massive
 fermions only.
With the help of identity~(\ref{p1}) it is very simple to derive
another relation, which is valid both for the massive and massless
cases~:
\begin{eqnarray}
 u(-k_1) \bar u(k_2) \Rightarrow
\frac{(m_1-\hat k_1)(1-\gamma^0) (1+\gamma^5)\gamma^{\xi}
(1+\gamma^0)(m_2 + \hat k_2) }
{4 \sqrt{2 (m_1+E_1)(m_2+E_2)}}
\end{eqnarray}
where $E_{1,2}$ are the fermion energies.

Identity (\ref{p1}) can be used for
the decomposition of the trace of an arbitrary $\gamma$-matrix
product into the sum of the products of traces. Indeed~:
\begin{eqnarray}
 && Tr(\gamma^{a_1} \gamma^{a_2} \cdots \gamma^{a_k} \cdots \gamma^{a_n})
\nonumber \\
 &&= \frac{1}{4} Tr(\gamma^{a_1} \gamma^{a_2} \cdots \gamma^{a_k}
(1+\gamma^5 + 1 - \gamma^5 )  \cdots \gamma^{a_n}
(1+\gamma^5 + 1 - \gamma^5 ))  \nonumber \\
 &&= \frac{1}{4}
 [ Tr(\gamma^{\lambda} \gamma^{a_1} \cdots \gamma^{a_k})
  Tr(\gamma^{\lambda} \gamma^{a_{k+1}}  \cdots \gamma^{a_n}) \nonumber \\
 &&-Tr(\gamma^5 \gamma^{\lambda} \gamma^{a_1} \cdots \gamma^{a_k})
  Tr(\gamma^5 \gamma^{\lambda} \gamma^{a_{k+1}} \cdots \gamma^{a_n})
 \nonumber \\
 &&+ Tr(\gamma^{\lambda} \gamma^{a_1} \cdots \gamma^{a_k} \gamma^{a_{k+1}})
  Tr(\gamma^{\lambda} \gamma^{a_{k+2}} \cdots \gamma^{a_n}) \nonumber \\
 &&-Tr(\gamma^5 \gamma^{\lambda} \gamma^{a_1} \cdots \gamma^{a_k}
 \gamma^{a_{k+1}})
  Tr(\gamma^5 \gamma^{\lambda} \gamma^{a_{k+2}} \cdots \gamma^{a_n}) ]
\label{p6}
\end{eqnarray}
One should note that in obtained relation (\ref{p6}) either
 two first or two last
lines survive, depending on the value of $n$ ($n = 2m$ or $n=4m$).

\section {}

Bearing in mind the evaluation of the amplitude expressions~(\ref{p5})
can be decomposed over the full set of a $\gamma$--matrixes
(when deriving this decomposition we take into account that
due to transition vertex $\bar b c \to B_c$ (see (\ref{gbc})) all Feynmann
 diagrams contain $\gamma^5$)~:
\begin{eqnarray}
&& -(m_1-\hat k_1) (1+\gamma^5)\gamma^{\xi} (m_2+\hat k_2) \gamma^5 =
  (m_1-\hat k_1) (1+\gamma^5)\gamma^{\xi} (m_2-\hat k_2) = \nonumber \\
&& = a_0^{\xi} + a_5^{\xi}\gamma^5 + V^{\xi \nu}\gamma^{\nu}
 + A^{\xi \nu} \gamma^5 \gamma^{\nu}
 + \frac{1}{2} T^{\xi \mu \nu} \gamma^{\mu} \gamma^{\nu} \label{b1}
\end{eqnarray}
\noindent Let us define two new vectors~:
\begin{eqnarray}
 && s^{\alpha} = m_2 k_1^{\alpha} + m_1 k_2^{\alpha} \quad \mbox{and}
 \quad u^{\alpha} = m_2 k_1^{\alpha} - m_1 k_2^{\alpha} \nonumber \\
 &&  (su) = 0, \quad s^2 = 2 m_1 m_2 (m_1 m_2 + (k_1 k_2)),
 \quad u^2 = 2 m_1 m_2 (m_1 m_2 - (k_1 k_2)). \nonumber
\end{eqnarray}
\noindent Then
\begin{eqnarray}
&& a_0^{\xi} =  -s^{\xi}, \quad
 a_5^{\xi}  =  u^{\xi}, \quad
 V^{\xi \nu}  =  \frac{1}{2m_1m_2} [ u^2 g^{\xi \nu}
 + s^{\xi} s^{\nu} - u^{\xi} u^{\nu}
 + i \varepsilon^{\xi \nu \lambda \sigma} s^{\lambda} u^{\sigma}],
 \nonumber \\
&& A^{\xi \nu} =  2m_1m_2 g^{\xi \nu} - V^{\xi \nu},
 \\
&& T^{\xi \mu \nu} = -T^{\xi \nu \mu}
 = g^{\xi \mu} u^{\nu} - g^{\xi \nu} u^{\mu}
 - i \varepsilon^{\xi \mu \nu \lambda} s^{\lambda}. \nonumber
\end{eqnarray}
Using decomposition (\ref{b1}) the
 calculation of the amplitude of subprocesses (\ref{r1}) and
(\ref{r2})  becomes very simple. The most comlicated expression contains
a trace of eight $\gamma$--matrixes. With the help of (\ref{p6})
it is very easy to evaluate the traces of products
(\ref{p5}) with $\gamma$--matrixes.
\begin{eqnarray}
 Q(A) \equiv Tr [\bar u(k_2) \gamma^5 A u(k_1)]  =
  \frac{1}{2 \sqrt{2 m_1 m_2}}
 Tr [(m_1-\hat k_1) (1+\gamma^5)\gamma^{\xi} (m_2 + \hat k_2) A].
\nonumber
\end{eqnarray}

Here we present the expressions for traces of $0,1, \cdots, 8$
$\gamma$--matrixes, needed for the evaluation of the amplitudes
(below we omit the coefficient $4/(2\sqrt{2 m_1 m_2})$).

Let us define a new tensor as follows~:
\begin{eqnarray*}
 S(\alpha \beta \mu \nu) \equiv
 g^{\alpha \beta} g^{\mu \nu} + g^{\alpha \nu} g^{\beta \mu} -
  g^{\alpha \mu} g^{\beta \nu}.
\end{eqnarray*}
Then one obtain~:
\begin{eqnarray*}
&&Q(1) = a_0^{\xi}; \quad  Q(\gamma^{\alpha}) = V^{\xi \alpha}; \quad
Q(\gamma^{\alpha} \gamma^{\beta}) = a_0^{\xi} g^{\alpha \beta}
- T^{\xi \alpha \beta}; \\
&&Q(\gamma^{\alpha} \gamma^{\beta} \gamma^{\delta}) =
V^{\lambda} S(\lambda \alpha \beta \delta) +
i \varepsilon^{\alpha \beta \delta \lambda} A^{\xi \lambda};  \\
&&Q(\gamma^{\alpha} \gamma^{\beta} \gamma^{\gamma} \gamma^{\delta}) =
a_0^{\xi} S(\alpha \beta \gamma \delta)
 - i a_5^{\xi} \varepsilon^{\alpha \beta \gamma \delta}  \\
&& - T^{\xi \alpha \beta} g^{\gamma \delta}
 + T^{\xi \alpha \gamma} g^{\beta \delta}
 - T^{\xi \alpha \delta} g^{\beta \gamma}
 - T^{\xi \beta  \gamma} g^{\alpha \delta}
 + T^{\xi \beta  \delta} g^{\alpha \gamma}
 - T^{\xi \gamma \delta} g^{\alpha \beta} ; \\
&&Q(\gamma^{\alpha} \gamma^{\beta} \gamma^{\gamma} \gamma^{\delta}
\gamma^{\mu}) = \\
&&V^{\xi \lambda}[ S(\sigma \lambda \alpha \beta )
 S(\sigma \gamma \delta \mu)
 + \varepsilon^{\sigma \lambda \alpha \beta}
 \varepsilon^{\sigma \gamma \delta \mu} ] \\
&& + i A^{\xi \lambda}[ \varepsilon^{\sigma \lambda \alpha \beta}
 S(\sigma \gamma \delta \mu)
 - S(\sigma \lambda \alpha \beta) \varepsilon^{\sigma \gamma \delta \mu}]; \\
&&Q(\gamma^{\alpha} \gamma^{\beta} \gamma^{\gamma} \gamma^{\delta}
\gamma^{\mu} \gamma^{\nu}) = \\
&&a_0^{\xi}[ S(\sigma \alpha \beta \gamma)
 S(\sigma \delta \mu \nu)
 + \varepsilon^{\sigma \alpha \beta \gamma}
 \varepsilon^{\sigma \delta \mu \nu} ]
 + i a_5^{\xi}[\varepsilon^{\sigma \alpha \beta \gamma}
 S(\sigma \delta \mu \nu)
 - S(\sigma \alpha \beta \gamma) \varepsilon^{\sigma \delta \mu \nu} ] \\
&& + T^{\xi \sigma \rho}[S(\sigma \delta \mu \nu)
       S(\rho \alpha \beta \gamma)
 + \varepsilon^{\sigma \delta \mu \nu}
\varepsilon^{\rho \alpha \beta \gamma}] \\
&& + \frac{1}{2}
 T^{\xi \sigma \rho} \varepsilon^{\lambda \eta \sigma \rho}
[ S(\eta \delta \mu \nu) \varepsilon^{\lambda \alpha \beta \gamma}
 - \varepsilon^{\eta \delta \mu \nu} S(\lambda \alpha \beta \gamma)]; \\
&&Q(\gamma^{\alpha} \gamma^{\beta} \gamma^{\gamma} \gamma^{\delta}
\gamma^{\mu} \gamma^{\nu} \gamma^{\sigma}) =  \\
&&V^{\xi \lambda} [S(\eta \nu \sigma \lambda)
 ( S(\rho \eta \alpha \beta) S(\rho \gamma \delta \mu)
 + \varepsilon^{\rho \eta \alpha \beta}
  \varepsilon^{\rho \gamma \delta \mu}) \\
 &&- \varepsilon^{\eta \nu \sigma \lambda}
 (\varepsilon^{\rho \eta \alpha \beta} S(\rho \gamma \delta \mu)
 - S(\rho \eta \alpha \beta) \varepsilon^{\rho \gamma \delta \mu}) ]  \\
&& + i A^{\xi \lambda} [\varepsilon^{\eta \nu \sigma \lambda}
 ( S(\rho \eta \alpha \beta) S(\rho \gamma \delta \mu)
+\varepsilon^{\rho \eta \alpha \beta} \varepsilon^{\rho \gamma \delta \mu}) \\
&&+ S(\eta \nu \sigma \lambda)
 (\varepsilon^{\rho \eta \alpha \beta} S(\rho \gamma \delta \mu)
 - S(\rho \eta \alpha \beta) \varepsilon^{\rho \gamma \delta \mu}) ]
\end{eqnarray*}

The above expressions for the traces
can be used for the evaluation of the amplitude of subprocesses
(\ref{r1}) and (\ref{r2}). For example, the expressions corresponding to
the diagram~$G.7$ (Fig.1) equal
\begin{eqnarray}
 M(G.7) = t^b t^a \sqrt{\frac{2}{m_1 m_2}} g_s^4
\frac{f_{B_c}}{12} (\frac{3}{4 (r_1k_2)})
 F_1^{\mu \nu} [-2(\epsilon_1 q_1) Q_1^{\mu\nu}
 + M Q_2^{\mu\nu} + Q_3^{\mu\nu}]
\end{eqnarray}
where $g_s$ is a strong coupling constant ($g^2_s = 4 \pi \alpha_s$);
$a$ and $b$ are the color indexes;
\begin{eqnarray*}
 F_1^{\mu \nu} = G(q_2)^{\mu \lambda} G(z_1)^{\nu \sigma}
 [2(\epsilon_2 q_2) g^{\lambda \sigma}
 + (2 r_2^{\lambda} - q_2^{\lambda}) \epsilon_2^{\sigma}
 - \epsilon_2^{\lambda} (r_2^{\sigma} + q_2^{\sigma}];
\end{eqnarray*}
here the $G(q)$ is the gluon propagator (see (\ref{prop}));
$q_2 = k_2 + \frac{m_2}{M}p$, $z_1 = q_1 - r_1$.

The tensors $Q_i^{\mu \nu}$ have the form~:
\begin{eqnarray*}
&& \gamma^{\mu}(M+\hat p)\gamma^{\nu} \Rightarrow Q_1^{\mu \nu} \\
&& = M(a_0^{\xi} g^{\mu \nu} - T^{\xi \mu \nu})
 - g^{\mu \nu}(V^{\xi \lambda} p^{\lambda})
 - i \varepsilon^{\mu \nu \lambda \sigma} p^{\lambda} A^{\xi \sigma}; \\
 &&\gamma^{\mu}\gamma^{\nu} \hat r_1 \hat \epsilon_1 \Rightarrow
 Q_2^{\mu \nu} \\
 &&= a_0^{\xi}  S_1^{\mu \nu} - i a_5^{\xi} \Delta_1^{\mu \nu}
  + T^{\xi \mu \lambda} S_1^{\nu \lambda}
  - T^{\xi \nu \lambda} S_1^{\mu \lambda}
  + g^{\mu \nu} (T^{\xi \alpha \beta} r_1^{\alpha} \epsilon_1^{\beta}) \\
 &&\gamma^{\mu} \hat p \gamma^{\nu} \hat r_1 \hat \epsilon_1 \Rightarrow
 Q_3^{\mu \nu} \\
 && = (p^{\lambda} V^{\xi \lambda}) S_1^{\mu \nu}
 + (p r_1) [- g^{\mu \nu}(\epsilon_1^{\lambda} V^{\xi \lambda})
 + \epsilon_1^{\mu} V^{\xi \nu} - \epsilon_1^{\nu} V^{\xi \mu}] \\
 &&+ i [  - g^{\mu \nu} \varepsilon^{\alpha \beta \lambda \sigma}
 r_1^{\alpha} \epsilon_1^{\beta} p^{\lambda} A^{\xi \sigma}
 - (p r_1) \varepsilon^{\mu \nu \lambda \sigma}
  \epsilon_1^{\lambda} A^{\xi \sigma} \\
 && +  A^{\xi \mu} \varepsilon^{\nu \alpha \lambda \sigma}
  p^{\alpha} r_1^{\lambda} \epsilon_1^{\sigma}
 -  A^{\xi \nu} \varepsilon^{\mu \alpha \lambda \sigma}
  p^{\alpha} r_1^{\lambda} \epsilon_1^{\sigma}].
\end{eqnarray*}
Where
\begin{eqnarray*}
S_1^{\mu \nu} = \epsilon_1^{\mu} r_1^{\nu} - \epsilon_1^{\nu} r_1^{\mu},
\quad \Delta_1^{\mu \nu} = \varepsilon^{\mu \nu \lambda \sigma}
r_1^{\lambda} \epsilon_1^{\sigma}.
\end{eqnarray*}

\newpage
\vspace{0.5cm}
\noindent
\underline{ {\bf Table 1.}} \\
The $B_c$--meson production cross section (without contribution from
 $B_c^{\ast}$, i.e. cross section  (\ref{s0}))
in $p \bar p$--collisions at $\sqrt{s} = 1.8$~TeV as the function of
evolution parameter $\sqrt{Q^2_{ev}}$, mass of a $b$ ($m_b$) and
$c$ ($m_c$) quarks ($M_{B_c} = 6.3$~GeV). The masses $m_b$ and $m_c$ are
in GeV, the cross section is in nb.
\begin{center}
\begin{tabular}{|c|c|c|c|}\hline
 $\sqrt{Q^2_{ev}}$  & $\sqrt{\hat s}$ & $\sqrt{\hat s}$/2 &
   $2 M_{B_c}$ \\ \hline
 $m_b = 4.5, \, \, m_c = 1.8$ & 15.2  & 24.5 & 28.2  \\ \hline
 $m_b = 4.8, \, \, m_c = 1.5$ & 19.0  & 30.8 & 32.8  \\ \hline
 $m_b = 5.0, \, \, m_c = 1.3$ & 31.1  & 52.4 & 56.0  \\ \hline
 $m_b = 5.2, \, \, m_c = 1.1$ & 28.4  & 46.4 & 51.8  \\ \hline
\end{tabular}
\end{center}

\newpage
{\bf Figure captions}
\vskip 0.5 cm
\begin{description}
\item{Fig.1} Some diagrams describing the subprocesses of quark--antiquark
annihilation (Q...) and gluon--gluon annihilation (G...) into
$B_c \, \bar b \, c$ final state.
\item{Fig.2} The diagram describing the transition of $b$ and $\bar c$
quarks into $B_c$--meson.
\item{Fig.3} The $B_c$--meson production cross section
 (without contribution from the $B_c^{\ast}$, see (\ref{s0}))
in $p \bar p$--collisions versus $\sqrt{s}$ (the parameters $\sqrt{Q^2_{ev}}$,
$m_b$ and $m_c$ are from (\ref{par})). The cross section is in nb,
 $\sqrt{s}$ is in GeV.
\item{Fig.4.a} The $B_c$--meson distribution over transverse
momentum in reaction (\ref{s0}). The $d\sigma/d p_{\bot}$ is in nb/GeV.
\item{Fig.4.b} The $b$--quarks (solid curve) and $c$--quarks (dashed
curve) distributions
over transverse momentum in reaction (\ref{s0}).
 The $d\sigma/d p_{\bot}$ is in nb/GeV.
\item{Fig.4.c} The $B_c$--meson distribution over  pseudorapidity
 in reaction (\ref{s0}). The $d\sigma/d \eta$ is in nb.
\item{Fig.4.d} The $b$--quarks (solid curve) and $c$--quarks
(dashed curve) distributions over pseudorapidity in reaction (\ref{s0}).
 The $d\sigma/d \eta$ is in nb.
\item{Fig.5} The distributions over transverse momentum (a)
and pseudorapidity (b) of the charged leptons from decay~(\ref{22})
in reaction~(\ref{s0}).  The dashed curve corresponds to leptons from
$J/\psi$ decay. $d\sigma/d p_{\bot}$ is in nb/GeV,
 $d\sigma/d \eta$ is in nb.
\item{Fig.6} The behaviour of integrated cross section (\ref{sig})
of the $B_c$--meson production with subsequent decay into charged leptons
 as a function of the minimum transverse momentum of lepton
 $p_{\top \, l}$.
\item{Fig.7} The distribution $d\Gamma_{B_c}/dM_{l \mu}$
 (normalized for the unit). The dashed curve is the distribution calculated
at $\chi = const$.
\item{Fig.8} The same distribution as on Fig.7, but over $M_{l \psi}$.
\item{Fig.9} The same distribution as on Fig.7, but over $M_{\top}$.

\end{description}

\end{document}